\documentclass[useAMS]{mn2e}
\usepackage{times}
\usepackage{rotating}
\long\def\symbolfootnote[#1]#2{\begingroup%
\def\thefootnote{\fnsymbol{footnote}}\footnote[#1]{#2}\endgroup}
\newcommand{\gae}{\lower 2pt \hbox{$\, \buildrel {\scriptstyle >}\over {\scriptstyle
\sim}\,$}}
\newcommand{\lae}{\lower 2pt \hbox{$\, \buildrel {\scriptstyle <}\over {\scriptstyle
\sim}\,$}}

\begin{document}

\title[Acceleration for GRB high-energy radiation]
{Implications of electron acceleration for high-energy radiation 
from gamma-ray bursts}

\author[Barniol Duran \& Kumar]{R. Barniol Duran$^{1,2}$\thanks
{E-mail: rbarniol@physics.utexas.edu, pk@astro.as.utexas.edu}
and P. Kumar$^{2}$\footnotemark[1] \\
$^{1}$Department of Physics, University of Texas at Austin, Austin, TX 78712, USA\\
$^{2}$Department of Astronomy, University of Texas at Austin, Austin,
TX 78712, USA}

\date{Accepted; Received; in original form 2010 March 30}

\pubyear{2010}

\maketitle

\begin{abstract}
In recent work we suggested that photons of energy $>$100 MeV detected from 
GRBs by the {\it Fermi} Satellite are produced via synchrotron emission in 
the external forward shock with a weak magnetic field -- consistent 
with shock compressed upstream magnetic field of a few tens of micro-Gauss.  
Here we investigate whether electrons can be accelerated to energies 
such that they radiate synchrotron photons with energy up to about 10 GeV 
in this particular scenario. We do this using two methods: (i) we check if 
these electrons can be confined to the shock front; and (ii) we calculate radiative 
losses while they are being accelerated.  We find that these electrons 
remain confined to the shock front, as long as the upstream magnetic field 
is $\gae 10 \mu$G, and don't suffer substantial radiative 
losses, the only condition required is that the external 
reverse shock emission be not too bright:
peak flux less than 1 Jy in order to produce photons of 100 MeV,
and less than $\sim$100 mJy for producing 1-GeV photons. We also find
that the acceleration time for electrons radiating at 100 MeV is a few seconds 
(in observer frame),
and the acceleration time is somewhat longer for electrons radiating
at a few GeV. This could explain the lack of $>$100 MeV photons
for the first few seconds after the trigger time for long GRBs reported
by the {\it Fermi} Satellite, and also the slight lag between photons of
GeV and 100 MeV energies. We model the onset of the external forward
shock light curve in this scenario and find it consistent with the
sharp rise observed in the 100-MeV light curve of GRB080916C and
similar bursts.

\end{abstract}


\section{Introduction}

The {\it Fermi} Satellite has detected 18 GRBs (Gamma-ray Burst) at energies $>$100 MeV by LAT 
(Large Area Telescope).  This emission can be described as follows.  The first 100 MeV 
photons arrive $\sim 1$ s (in the host galaxy frame) after the trigger time, 
for long GRBs; the trigger time is 
the time when low energy photons ($\sim1$ MeV) are first detected by the GBM (Gamma-ray Burst 
Monitor aboard {\it Fermi}). The 100 MeV light curve rises fast until it peaks and then it 
decays as a single power-law for a long duration of time (of order 10$^3$ s) -- much 
longer than the duration of the lower energy photons detected by GBM -- until it falls 
below the detector's sensitivity. Radiation above 100 MeV from GRBs has been suggested to be produced 
via the synchrotron mechanism 
in the external forward shock (Kumar \& Barniol Duran 2009, 2010); the external forward shock 
scenario was first proposed by Rees \& M\'esz\'aros (1992), M\'esz\'aros \& Rees (1993),
Paczy\'nski \& Rhoads (1993), and since then it has 
been used widely, see, e.g., M\'esz\'aros \& Rees (1997), Sari, Piran, Narayan (1998), 
Dermer \& Mitman (1999), for a comprehensive review see, e.g., Piran (2004) and references therein.  After 
our initial suggestion, many groups have also considered and provided evidence for this 
origin of the $>$100 MeV radiation (Gao et al. 2009; Corsi, Guetta, Piro 2010; De Pasquale et al. 2010;
Ghirlanda, Ghisellini, Nava 2010; Ghisellini, Ghirlanda, Nava 2010).  The magnetic field required
for this model is consistent with being  produced via shock-compressed seed magnetic field 
in the CSM (circum-stellar medium) of strength of a few tens of micro-Gauss.  The peak of the 
100 MeV light curve can be attributed to the deceleration time which is the time it 
takes for the GRB-jet to transfer about half of its energy to the external medium.

We investigate in this work whether electrons in the external forward shock can be
accelerated to sufficiently high Lorentz factors, even for a 
small CSM magnetic field of a few tens of $\mu$G, so that the
synchrotron radiation can extend to $\sim 10$ GeV as seen by {\it Fermi}/LAT for a number
of GRBs.  We study the electrons acceleration in the context of 
diffusive shock acceleration (e.g., Krymskii 1977, Axford, Leer \& Skadron 1978, 
Bell 1978, Blandford \& Ostriker 1978, 
Blandford \& Eichler 1987), which was developed for non-relativistic shocks and has now been 
developed to consider relativistic shocks (semi-) analytically 
(e.g. Gallant \& Achterberg 1999, Achterberg et al. 2001) and recently 
using 2D particle-in-cell simulations (e.g. Spitkovsky
2008a,b, Keshet et al. 2009).  We assume that the electrons acceleration 
proceeds in the Bohm diffusion limit and that the magnetic field downstream 
is simply shock-compressed upstream magnetic field (other possibilities
are considered in, e.g., Milosavljevi\'c \& Nakar 2006, Sironi \& Goodman 2007, 
Goodman \& MacFadyen 2008, Couch, Milosavljevi\'c, Nakar 2008). 

If the downstream magnetic field is simply the shock-compressed 
large-scale upstream field, then the field component perpendicular to the shock 
normal is amplified, while the parallel component is not.  In this case,  
the downstream magnetic field will be mainly pointing to the direction 
perpendicular to the shock front normal, therefore particles trying to cross the shock front 
from downstream to upstream will find it difficult to catch up with the shock front, 
which moves with a speed of $\sim c/3$ with respect to the downstream medium 
(see, e.g., Achterberg et al. 2001, Lemoine, Pelletier \& Revenu 2006, 
Pelletier, Lemoine \& Marcowith 2009).  One way that the 
particles might return to the upstream is if there is efficient cross-field 
diffusion of particles, which might occur if turbulent magnetic field is produced 
downstream (Jokipii 1987, Achterberg \& Ball 1994, Achterberg et al. 2001).  
In principle, the turbulent magnetic field could dominate the shock-compressed field 
throughout the downstream region.  However, it seems that although some 
turbulence is present just downstream of the shock front it does not persist across the entire downstream region 
(see recent simulations by Sironi \& Spitkovsky 2010 that show that magnetic field
is amplified only right behind the shock front and returns to the shock-compressed value
far downstream). In this case, much of the radiation is produced by particles swept 
downstream where the turbulence has died out and the magnetic field is consistent with the 
shock-compressed value. We also note that as long as the thickness
of the turbulent magnetic field layer is smaller than the thickness
of the shocked fluid divided by $(B_t/B_d)^2$ then the energy loss in
the turbulent layer is small; $B_t$ is the turbulent magnetic field
strength and $B_d$ is the shock-compressed magnetic field.  Therefore, 
in this work we neglect energy loss in the turbulent magnetic field layer since it  
persists for a very short distance compared to the thickness of the shocked 
fluid (see, e.g., Keshet et al. 2009 and references therein).

This work is organized as follows.  In Section 2 we address the question of high-energy electron
confinement upstream and downstream of the shock front, and also 
radiative losses suffered by electrons in between acceleration.
Also, in Section 2, we discuss the lag of the $>$100 MeV
light curves observed by {\it Fermi} LAT for several GRBs in light of our results on
electron acceleration. In Section 3, we calculate the rise of the external forward shock light curve, 
taking into consideration the non-zero time to accelerate electrons to high enough 
energies so they can radiate at $>$100 MeV. We present our conclusions in 
Section 4.

\section{Electron acceleration for $>$100 MeV emission}

\subsection{Electron confinement}

It is widely believed that electrons in non-relativistic shocks undergo diffusive 
shock acceleration. (e.g., Krymskii 1977, Axford et al. 1978, Bell 1978, 
Blandford \& Ostriker 1978, Blandford \& Eichler 1987). In the context of 
relativistic shocks, it has been shown that electrons gain 
energy each time they cross the
shock front by a factor of $\sim 2$, except on the first crossing when they 
gain energy by a factor of the Lorentz Factor (LF) of the shock front (Achterberg et al. 2001).  

In order for electrons to turnaround while up/down stream and cross the shock front,
their Larmor radius should be smaller than the size of the system, i.e. electrons
should be confined to the system in order to be accelerated. In this subsection,
we explore the confinement of electrons in the external forward shock model when
the magnetic field in the unshocked medium, upstream of the shock front, is a 
few tens of $\mu$G in strength, and the magnetic field in the shocked 
medium, downstream of the shock front, is simply the shock-compressed upstream 
field.

The highest photon energy detected for {\it Fermi} GRBs is on the order of 10 GeV. 
We first calculate the random LF in the downstream co-moving frame, 
$\gamma_e$, of electrons radiating 10 GeV photons via synchrotron radiation, 
because these electrons have the largest Larmor radius and thus give us stricter 
confinement requirements.  The synchrotron frequency in observer frame is 
$\nu_{syn} =  e B_d \gamma_e^2 \Gamma / 2 \pi m_e c (1+z)$, where $\Gamma$ is the bulk 
LF of the shocked fluid measured in the upstream rest frame (lab frame), 
$B_{d}$ is the magnetic field downstream (measured in the local rest frame), 
$z$ is the redshift, $m_e$ and $e$ are the electron's mass and charge, respectively, 
and $c$ is the speed of light (Rybicki \& Lightman 1979).  We convert the synchrotron frequency to 10 GeV, i.e., 
$\nu_{10} = h \nu_{syn}/1.6 \times 10^{-2}$ erg, where $h$ is the Planck constant and 
10 GeV corresponds to $1.6 \times 10^{-2}$ erg. Using the convention 
$Q_{x}=Q/10^{x}$ and solving the last expression for $\gamma_e$ yields 

\begin{equation}
\gamma_e = 1.5\times10^8 \nu_{10}^{1/2}(1+z)^{1/2}\Gamma_3^{-1} B_{u,-5}^{-1/2},
\end{equation} 
where $B_{u}$ is the magnetic field upstream, which is the magnetic field in the CSM.   
To obtain (1) we have assumed that the magnetic field in the downstream region 
is $B_d = 4 \Gamma B_{u}$ (Gallant \& Achterberg 1999, Achterberg et al. 2001; note 
that the shock front LF measured in the lab frame is $\Gamma_s = \sqrt{2} \Gamma$, Blandford \& McKee 1976), 
i.e. $B_d$ is the shock-compressed magnetic field in the upstream (lab frame), 
which is what we have found for {\it Fermi} GRBs (Kumar 
\& Barniol Duran 2009, 2010).

The electrons' LF in the rest frame of the upstream plasma is  $\gamma_e \Gamma$, therefore, 
the Larmor radius in the upstream is given by 
\begin{equation}
R_{L,u}=\frac{m_e c^2 \gamma_e \Gamma}{eB_{u}}=({2.6\times10^{19} \rm cm})\, 
\nu_{10}^{1/2}(1+z)^{1/2} B_{u,-5}^{-3/2}, 
\end{equation} 
where we made use of (1) to eliminate $\gamma_e$. Comparing the Larmor 
radius with the size of the system upstream, $R$, which is given by the blast wave 
radius in the host galaxy rest frame --- $R = 2c\Gamma^2 t/(1+z) \sim10^{17}$cm
(where $t\sim$ a few seconds and $\Gamma\sim10^3$ is the blast wave Lorentz factor, e.g.
Abdo et al. 2009a) --- we find 
that $R_{L,u} \gg R$. This might suggest that electrons of $\gamma_e\sim10^8$ are not 
confined to the system. However, an electron upstream of the shock front travels 
only a distance $\sim R_{L,u}/\Gamma$ before returning to the downstream,  
because by the time the angle between electrons' velocity 
vector and the normal to the shock front exceeds $\sim 1/\Gamma$, the shock front 
catches up with the electron and sweeps it back downstream (Achterberg et al. 2001).
Therefore, for electron confinement upstream one should compare $R_{L,u}/\Gamma$ with $R$:
\begin{equation}
\frac{R_{L,u}}{\Gamma R}=0.26\,{\nu_{10}^{1/2}(1+z)^{1/2}\over 
   \Gamma_3 B_{u,-5}^{3/2} R_{17}} =
1.1 \frac{\nu_{10}^{1/2}   t_3^{1/8} (1+z)^{3/8}}{B_{u,-5}^{3/2} (E_{54}/n_0)^{3/8} },
\end{equation}
where $E$ is the energy in the blast wave, $t$ is the time since the burst trigger 
in observer frame, and $n$ is the number density of particles in the CSM; 
in deriving the second equality we made use of the time dependence of $\Gamma$ and $R$
in the external forward shock scenario for a homogeneous CSM (Sari, Piran, Narayan 1998).
For $B_{u,-5}\approx4$ found for the {\it Fermi} bursts,
$R_{L,u}/(\Gamma R)\lae 0.2$, and, thus, electrons radiating at 10 GeV cannot escape from
the upstream side of the shock front; note that this conclusion holds for at least
several hours in the observer frame.

One should also check for electron confinement downstream. Here, the Larmor radius is 
smaller than it is upstream, because the magnetic field is larger 
by at least a factor of $4\Gamma$ due to shock compression. Therefore, the requirement for 
the confinement of electrons downstream is automatically satisfied whenever it is satisfied
upstream. 

We conclude that there is no problem confining external forward shock electrons that 
radiate $\sim$10 GeV synchrotron photons by the CSM magnetic field
of strength $\gae10$ $\mu$G. 

\subsection{Radiative losses during electron acceleration}

Electrons suffer radiative losses while being accelerated that could prevent them from reaching 
LFs of $\sim10^8$ that are needed for radiating photons of 10 GeV via the synchrotron process.
In this section, we ascertain whether or not the radiative losses suffered by electrons -- 
due to synchrotron and inverse-Compton processes -- are small compared with the 
energy gain in each round of crossing the shock front. We do this by comparing the total 
radiative cooling time-scale, $t'_{cool}$, which is the time-scale for electrons to lose 
half of their energy, with the acceleration time-scale.   

For the case of ultra-relativistic shocks when the downstream magnetic 
field is simply the shock compressed upstream field, the upstream and 
downstream residency times for electrons are approximately equal, 
when particle diffusion is in the Bohm limit (Gallant \& Achterberg 1999, Achterberg et al. 2001). 
Thus, the time it takes for electrons to make one complete cycle across the 
shock front is about twice the upstream residency time, and the upstream
residency time is on the order of the gyro-time in the shock front co-moving frame (Baring 2004).
In the lab frame, their upstream residency time is on the order of the time it takes them to travel a distance
$\sim R_{L,u}/\Gamma$. Since the Larmor radius ($R_{L,u}$) increases with increasing 
electron energy, the last shock crossing dominates the total upstream residency 
time. Thus, the time, in the co-moving frame of the blast wave, that electrons spend 
during the last cycle of crossing the shock front (upstream $\rightarrow$ downstream 
$\rightarrow$ upstream) before getting accelerated to Lorentz factor $\gamma_e$ 
-- given by (1) --is:

\begin{equation}
t'_{s}\sim \frac{2 R_{L,u}}{c\Gamma^2}= ({1.7\times10^3 \rm s})\, \nu_{10}^{1/2}(1+z)^{1/2}
    \Gamma_3^{-2} B_{u,-5}^{-3/2}.
\end{equation} 

Taking into account the energy loss that these electrons experience because 
of radiative cooling, the acceleration time-scale, in the blast wave co-moving frame, 
is given by
\begin{equation}
 t'_{acc}(\gamma_e) \approx t'_{eq}(\gamma_e) +  t'_s(\gamma_e),
\end{equation}
where $t'_{eq}(\gamma_e)$ is the elapsed time since the beginning of the
explosion when $t'_s(\gamma_e) = t'_{cool}(\gamma_e)/2$ (shock front crossing 
time should be equal to at least half of the the radiative cooling time in 
order to reach a particular $\gamma_e$).  At $t'_{eq}$ the electron barely 
reaches $\gamma_e$, therefore, it needs an extra time on the order of $\sim t'_s$ to fully reach 
the desired $\gamma_e$.  If $t'_s > t'_{cool}/2$, then the radiative 
cooling is too strong and prevents the electron from reaching the desired $\gamma_e$. 
In the sub-sections below we discuss synchrotron and inverse-Compton losses
and calculate the radiative cooling time.

\subsubsection{Synchrotron losses}       

The synchrotron cooling time-scale (in the blast wave co-moving frame) in the upstream of the
shock front is $t'_{syn,u}=6 \pi m_e c/\sigma_T B_{u}^2 \gamma_e \Gamma^2$, where $\sigma_T$ 
is the Thomson scattering cross-section. We find that the synchrotron cooling time for an 
external froward shock electron with LF given by (1) is 
\begin{equation}
t'_{syn,u}=({5.2\times10^{4} \rm s})\, \nu_{10}^{-1/2}(1+z)^{-1/2} \Gamma_3^{-1} B_{u,-5}^{-3/2}.
\end{equation}
Since $t'_{syn,u}>t'_{s}$ by a factor of 30, then synchrotron cooling in the upstream is 
unimportant for electrons radiating at 10 GeV.  

Next, we calculate synchrotron losses in the downstream. Since $t'_{syn} \propto B^{-2}$, the 
synchrotron loss rate is larger downstream because of the larger magnetic field. 
For shock-compressed magnetic field downstream, $B$ is larger than upstream field by
a factor 4 (in the blast wave co-moving frame), and therefore $t'_{syn,d}=t'_{syn,u}/16$.
The effective synchrotron cooling time for electrons of LF
given in (1) is $t'_{syn}\approx[1/2 t'_{syn,d} + 1/2 t'_{syn,u}]^{-1}$, which gives

\begin{equation}
t'_{syn} = (6.1\times10^{3}{\rm s})\, \nu_{10}^{-1/2}(1+z)^{-1/2} \Gamma_3^{-1} B_{u,-5}^{-3/2}.
\end{equation} 

We see from (4) that $t'_{syn} \sim 4 t'_{s}$ for electrons that produce synchrotron 
photons of 10 GeV energy, and therefore the maximum synchrotron photon energy --- obtained by 
setting $t'_{s}=t'_{syn}$ --- is $\nu_{max, syn}\sim 40 \Gamma_3 (1+z)^{-1}$ GeV 
(see, e.g., Guilbert, Fabian, Rees 1983, de Jager et al. 1996, Cheng \& Wei 1996).

\subsubsection{Inverse-Compton losses}

In this sub-section we calculate the inverse-Compton (IC) cooling time-scale for electrons. 
The inverse-Compton cooling 
time depends on the energy density of photons, and on the electron LF. Electrons in 
the external forward shock region are exposed to photons from three different
sources of radiation: (a) prompt $\sim$MeV $\gamma$-ray radiation which carries
most of the energy release in GRBs; (b) synchrotron radiation produced in 
the external forward shock heated CSM and (c) radiation produced in the external 
reverse shock heated GRB-jet. We will consider all of these sources
in our estimate for the IC cooling time. All calculations will be carried out in 
the rest frame of the shocked CSM.

The IC cooling time is given by
\begin{equation}
t'_{IC} =\frac{3 m_e c^2}{4 \int d\nu\,\sigma F'(\nu) \gamma_e},
\end{equation} 
where $F'(\nu)$ is the energy flux in radiation per unit frequency in the co-moving frame of the shocked CSM,
$\nu$ is photon frequency in observer frame, and $\sigma$ is the cross-section for 
interaction between electrons and photons;
$\sigma\approx \sigma_T$ (Thomson cross-section) when $\nu< \Gamma m_e c^2/[(1+z)
h \gamma_e]\equiv \nu_{kn}$, and for $\nu\gg\nu_{kn}$, $\sigma\approx\sigma_T(\nu/\nu_{kn})^{-1}$.
Thus, an approximate equation for the IC cooling time is

\begin{equation}
t'_{IC} \approx {3 m_e c^2\over 4 \sigma_T \gamma_e}\left[ F'(< \nu_{kn}) +
     {\nu_{kn}\over\nu_p} F'(> \nu_{kn}) \right]^{-1},
\end{equation}
where $ F'(< \nu_{kn})$ is photon energy flux in the shock co-moving frame
below the frequency $\nu_{kn}$ and $F'(>\nu_{kn})$ is the flux above $\nu_{kn}$.
The frequency at the peak of the $\nu F(\nu)$ spectrum is $\nu_p$ (in observer frame, 
i.e., co-moving synchrotron peak frequency boosted by a factor of $\Gamma$
and redshift corrected)
and $\nu_{kn}$, the Klein-Nishina frequency in the observer frame, for an electron of 
LF $\gamma_e$, is $h \nu_{kn} \approx ({\rm 5\, eV})\; \Gamma_3 \gamma_{e,8}^{-1} (1+z)^{-1}$.  
We note that for $\nu_{kn}\gae\nu_p$, only the first term in (9) should
be kept.

The co-moving energy flux in radiation is related to the observed 
bolometric luminosity by:
\begin{equation}
F'(\lae\nu_p) \sim {L_{obs}\over 4\pi R^2 \Gamma^2}.
\end{equation}
Combining (9) and (10) we find
\begin{eqnarray}
t'_{IC} \sim \frac{3 \pi R^2 \Gamma^2 m_e c^2}{\sigma_T L_{obs} \gamma_e}
   \left(\frac{\nu_{kn}}{\nu_p}\right)^{-1} \left[1+\left({\nu_{kn}\over\nu_p}
   \right)^\alpha \right]^{-1},
\end{eqnarray}  
where $\alpha$ is the spectral index, i.e. $F'(\nu)\propto \nu^\alpha$ for
$\nu_{kn}<\nu<\nu_p$; for $\alpha>0$ the term in the square bracket is of
order unity. The above equation is valid only when $\nu_{kn} < \nu_p$.

As mentioned before, there are three different sources of photons that interact 
with electrons in the the external forward shock. We analyze these cases separately.

\noindent{\bf Case (a)}: The prompt $\gamma$-ray emission in GRBs -- the origin of which is
still uncertain -- often has a low energy spectral index $\alpha\sim 0$, and the
spectrum peaks at $\nu_{p,6}\sim1$. The luminosity of this component is the highest
of the three cases considered; $L_{obs,52} \sim 10$ for {\it Fermi} GRBs that have $>$10$^2$MeV 
emission. The cooling time, obtained from (11), for this case is
\begin{equation}
t'_{IC,a}=({2.2\times10^3 \rm s}\,)\frac{R_{17}^2 \Gamma_3 \nu_{p,6}(1+z)}{L_{obs,53}} 
\left[1+\left({\nu_{kn}\over\nu_p}\right)^\alpha \right]^{-1}
\end{equation}  

\noindent{\bf Case (b)}: The external forward shock synchrotron spectrum peaks at $\sim 100$keV 
(before the deceleration time), and the spectral index between $\nu_{kn}$ and $\nu_p$ is
$\alpha\sim 1/3$. The luminosity from the external forward shock 
is $L_{obs,52} \sim 0.1$ at the deceleration radius ($R_d$),
and at smaller radius it decreases as $\sim R^3$.  Therefore, we find  
from (11) that, for $R \le R_d$, 

\begin{equation}
t'_{IC,b}\approx ({2.2\times10^4 \rm s}\,) R_{17}^{-1} \Gamma_3 \nu_{p,5} L_{obs,51}^{-1} R_{d,17}^3 (1+z).
%    \quad\quad {\rm for\;} R \le R_d.
\end{equation} 

\noindent{\bf Case (c)}: If the GRB-jet is composed
of protons and electrons, then the interaction of the jet with the CSM will heat 
up these particles by the reverse shock propagating into the cold jet, and the 
synchrotron radiation produced would be very effective at cooling electrons
in the forward shock region. This is because the peak of the reverse shock emission
at the deceleration time is typically at a few eV (Sari \& Piran 1999a), 
which is of order $\nu_{kn}$ for electrons of $\gamma_e \sim 10^8$.  
Since $\nu_p \sim \nu_{kn}$, then we can keep only the first term in (9), 
and use (10) for flux in the calculation of the cooling time.
The observed luminosity (at the deceleration time) is given by 
$L_{obs,d} \approx 4 \pi d_L^2 \nu_{p,d} F_{p,d}$, where $d_L$ is the 
luminosity distance, and $F_{p,d}$ and $\nu_{p,d}$ are the observed external 
reverse shock flux and peak energy at the deceleration time, respectively.
Thus, the IC cooling time-scale for electrons in the external forward shock region, due to 
the radiation produced in the reverse shock heated GRB-jet, is

\begin{equation}
t'_{IC,c} \approx ({400 \rm s})\, R_{d,17}^2 \Gamma_{d,3}^2 d_{L,28}^{-2} \gamma_{e,8}^{-1} 
   \nu_{p,d}^{-1} F_{p,d}^{-1} ,
\end{equation} 
where $\Gamma_d$ is the LF of the GRB-jet at the deceleration time, 
$F_{p,d}$ is in Jansky (Jy) and $\nu_{p,d}$ in eV; the reverse peak flux can be $\sim 1$ Jy for very bright
bursts such as GRB 990123 (Sari \& Piran 1999b). 

%\begin{equation}
%t'_{IC,c} \approx ({400 \rm s})\, R_{17}^{-y} \Gamma_{d,3}^2 d_{L,28}^{-2} \gamma_{e,8}^{-1} 
%   \nu_{p,d}^{-1} F_{p,d}^{-1} R_{d,17}^{2+y}
%   \quad\quad {\rm for\;} R \le R_d
%\end{equation}
%In the external reverse shock scenario, before the deceleration time, 
%all the parameters in equation (11) change with time. Their time dependencies can be 
%found in table 2 of McMahon, Kumar, Piran (2006). Using the information in that table, 
%we find that $F'(\lae\nu_p)\propto R^{y}$, where $y = [1.3, -3.7, 7.2, 2]$, for the cases of 
%thick shell, $s=0$ and $s=2$, and thin shell, $s=0$ and $s=2$, respectively 
%(the CSM density fall-off is taken to be $\propto R^{-s}$ and the power-law index for 
%injected electron energy distribution is $p=2.4$).

%For $\nu_p=2$, and $d_{L,28}=4.9$, which corresponds to a redshift of $z=2$, we find
%\begin{equation}
%t'_{IC,c} \approx ({8 \rm s})\, R_{17}^2 \Gamma_3^2 \gamma_{e,8}^{-1} F_p^{-1}.
%\end{equation}  

The total IC cooling time is 

\begin{equation}
t'_{IC} = \left[\frac{1}{t'_{IC,a}}+\frac{1}{t'_{IC,b}}+\frac{1}{t'_{IC,c}}\right]^{-1}
\end{equation}
and, finally, the total radiative cooling time, $t'_{cool}$, is given by

\begin{equation}
t'_{cool} = \left[\frac{1}{t'_{syn}} + \frac{1}{t'_{IC}}\right]^{-1}.
\end{equation}

%This cooling time-scale should be compared with the shock crossing time of the 
%electrons, $t'_{s}$, given in (4) to determine if electrons can be accelerated
%to some desired LF $\gamma_e$ or not.

\subsection{Application to {\it Fermi} GRBs}

In this section, we analyze 3 GRBs detected by {\it Fermi}: GRB080916C, 090510 and 090902B 
(Abdo et al. 2009a,b,c).  The relevant data for each burst are tabulated in Table 1.
We apply the above general results to these three GRBs, and determine
the time it would take for electrons in the external forward shock for these bursts to be 
accelerated -- via shock acceleration -- to LFs capable of producing synchrotron 
photons of energies 100 MeV and 1 GeV (Table 2).

The external reverse shock emission depends on the highly uncertain magnetic field
strength in the GRB-jet, and it is therefore difficult to estimate with any 
confidence. We calculate $t'_{cool}$ by neglecting the contribution of 
reverse shock emission to inverse-Compton cooling of electrons ($t'_{IC,c}$), and this
provides a lower bound to $t'_{acc}$ which is reported in Table 2
as a fraction of the deceleration time, $t'_d=({1.7\times10^3 \rm s})\, R_{d,17} \Gamma_{d,3}^{-1}$, 
for several {\it Fermi} bursts. We also provide in Table 2 an
upper limit for the external reverse shock peak flux that is obtained by the condition
that $t'_{cool} = t'_s$, at the deceleration time, when
the contribution of the external reverse shock emission is included in 
the calculation of $t'_{cool}$.

%%%%%%%%%%%%%%%%%%%%%%%%%%%%%%Begin The GRB Table 1 %%%%%%%%%%%%%%%%%%%%%%%%%%%%%
\begin{table}
%\begin{minipage}{\textwidth}
\begin{center}
\begin{tabular}{cccccc}
\hline
GRB & $\alpha$ & $\nu_{p} [MeV]$ & $B_{u,-5}$  & $L_{obs,53}^{prompt}$ & $L_{obs,51}^{ES}$  \\
\hline 
080916C  & 0  & 0.5 & 4 & $1.5$ & $30$\\
%\hline
090510 & 0.4 & 2.8 & 2 & $3.6$ & $1.6$\\
%\hline
090902B & 0.4 & 0.7 & 2 & $1.2$ & $1.7$\\
\hline
\end{tabular}
\end{center}
%\end{minipage}
\caption{{\small The main quantities used in our analysis for three {\it Fermi} GRBs. 
$\alpha$ is the approximate spectral energy index, during the prompt emission phase,
 below the peak of the spectrum
($f_\nu\propto \nu^{\alpha}$ for $\nu<\nu_p$), $\nu_{p}$ is the 
observed peak of the spectrum; $B_{u,-5}$ is the average upstream magnetic 
field, in units of 10 $\mu$Gauss, obtained by modeling of the data for these 
bursts (Kumar \& Barniol Duran 2010);
$L_{obs,53}^{prompt}$ and $L_{obs,51}^{ES}$ are the approximate observed 
isotropic equivalent luminosities of the prompt $\gamma$-rays and 
external forward shock emission at the deceleration time, respectively. 
Data are taken from Abdo et al. (2009a,b,c).
$B_u$ was obtained by setting three simple constraints while modeling
the external forward shock emission: 1. Its flux at 100 MeV should agree with 
the observed value, 2. Its cooling frequency should be below 100 MeV at early 
times for consistency with the observed spectrum, and 3. Its flux at 100 keV
should be smaller than the observed value during the observed steep decay, 
so that the external forward shock emission doesn't prevent 
the 100 keV to decay steeply (Kumar \& Barniol Duran 2009, 2010).}}
\end{table}
%%%%%%%%%%%%%%%%%%%%%End the GRB Table 1  %%%%%%%%%%%%%%%%%%%%%%%%%%%%%%%%

\noindent{\it GRB080916C}: The first $>$100 MeV photons arrived $\sim 3$s 
after the trigger time and then the 
100 MeV light curve rose rapidly, as $\sim t^6$, and peaked at 
$\sim 5$s (Abdo et al. 2009a).  After the peak, which 
we identify as the deceleration time, $t_d$, the flux decayed as a 
single power-law (this power-law is consistent with the expectation of the 
external forward shock model). So the first $>$100 MeV photons arrived 
at $t/t_d \sim 0.6$, and photons of energies $>$GeV were detected at 
$\sim 7$ s ($t/t_d \sim 1.4$). The highest energy photon, $\sim 13$ GeV,
was detected $\sim 16$ s after the trigger time ($t/t_d \sim 3$).

For electrons to produce 100 MeV photons their LF should be $\sim 10^7$ 
for this burst, and for 1 GeV photons the required $\gamma_e>3 \times 10^7$; 
we used $B_{u,-5}\sim4$ as suggested by the data for this burst (Kumar \& Barniol Duran
2009) -- see Table 1. The acceleration time for electrons to attain these LFs is calculated using
(5); note that our 
theoretical estimates are roughly consistent with the observed time-scales 
for GRB080916C to within a factor $\sim2$ uncertainty of our estimates
(Table 2).

\noindent{\it GRB090510}: For GRB090510 (Abdo et al. 2009b) 
there was a short delay in the detection 
of $>$100 MeV photons by $\sim 0.1$ s (we take the trigger time to be 
$\sim 0.5$ s after the GBM trigger, because of the presence of a precursor).  
The 100 MeV light curve peaked at $\sim 0.2$ s (which we associate with the
deceleration time), and so the arrival of the first $>$100 MeV photons was at 
$t/t_d \sim 0.5$. Higher energy photons arrived later: $>$1 GeV photons started
 arriving at $t_d$, and $\sim 10$ GeV photons arrived slightly 
after $t_d$. As shown in Table 2 these results are roughly consistent with our
estimates within a factor of 2.

\noindent{\it GRB090902B}: The 100 MeV light curve for 
this burst peaked at $\sim 10$s, which we identify as $t_d$, 
and the first $>$100 MeV photons were detected at $\sim 3$s after the trigger time 
(Abdo et al. 2009c), i.e.  $t/t_d \sim 0.3$.  Most of the GeV photons 
arrived at $\sim t_d$. The first 10 GeV photon is detected at $\sim 12$ s.
The highest energy photon detected was $\sim 30$ GeV at 80 s, i.e. at $\sim 8 t_d$
\footnote{At this time, the LF has dropped by a factor of $8^{3/8}\sim 2$, and at $z=1.8$, 
$\nu_{max,syn} \sim 10$ GeV, a factor of $\sim 4$ smaller than the observed value. 
It can be shown that inhomogeneous magnetic fields lead to an increase
of $\nu_{max,syn}$ by about an order of magnitude.}. The arrival time for the
first $>$100 MeV photons from this burst agrees with the electron 
acceleration time (Table 2). 

To summarize the main results of this section, it takes a few seconds for electrons
in the external forward shock to be accelerated to a LF so that they can produce 
100 MeV photons,
and it takes a bit longer time for them to produce GeV photons. For this reason, 
GeV photons lag the 100 MeV radiation. If the external reverse shock flux
is high ($\sim 1$ Jy), then the first 100 MeV photons will be detected after
the deceleration time, and 10 GeV photons will be 
detected much later ($\sim 10t_d$), when the reverse shock flux has decreased substantially . 
If the external reverse shock flux is 
small ($\sim 10$ mJy), then the first 100 MeV photons will arrive at about a 
third of the deceleration time, and GeV photons will be 
detected starting from close to the deceleration time.

%%%%%%%%%%%%%%%%%%%%%%%%%%%%%%Begin The GRB Table 2 %%%%%%%%%%%%%%%%%%%%%%%%%%%%%
\begin{table}
%\begin{minipage}{\textwidth}
\begin{center}
\begin{tabular}{ccccc}
\hline
    &       & Expected      & Observed & \\ 
GRB & $\nu$ & $t_{acc}/t_d$ & $t/t_d$ & $F_{p,d}^{max}$\\
\hline 
080916C  & 100 MeV  & 0.3 & 0.6 & 0.30 \\
         & 1 GeV    & 0.6 & 1.4 & 0.02 \\
\hline
090510   & 100 MeV  & 0.3 & 0.5 & 9.90 \\
         & 1 GeV    & 0.6 & 1 & 0.90 \\
\hline
090902B  & 100 MeV  & 0.3 & 0.3 & 1.20 \\
         & 1 GeV    & 0.7 & 1 & 0.10 \\
\hline
\end{tabular}
\end{center}
%\end{minipage}
\caption{{\small $t_{acc}/t_d$ is the ratio of
the time for electron acceleration to a specific energy (corresponding to
synchrotron frequency given in column 2) and the deceleration time; it
is a measure of the delay, with respect of the trigger time,
for photons of a given energy to arrive at the observer when the 
external reverse shock emission is smaller than given in the last 
column of the Table.  The observed time delay of photons in column 2 
with respect to the trigger time is $t/t_d$. $F_{p,d}^{max}$ (in Jy) is the maximum possible observed 
external reverse shock peak flux, so that electrons can be accelerated 
to produce photons of energy given in column 2 at $t_d$. }}
\end{table}
%%%%%%%%%%%%%%%%%%%%%End the GRB Table 2  %%%%%%%%%%%%%%%%%%%%%%%%%%%%%%%%

\section{Steep rise of the high-energy photon light curve}

In this section, we calculate the onset of light curves of high-energy photons ($>$100 MeV).
According to the standard external forward shock model, and assuming instantaneous
acceleration of electrons, the observed flux rises as $t^3$ when the CSM-density is
homogeneous. We show here that the
light curve rises much more steeply -- similar to what is seen by {\it Fermi}/LAT data -- 
when finite time for electron acceleration is taken into consideration, as was
suggested by Kumar \& Barniol Duran (2009). 

We calculate the rise of the light curve using a simple model.
The external shock emission at some frequency $\nu$ is zero until the
blast wave reaches a radius $R_0$, which is set by the 
time-scale for electrons to be accelerated to a LF so that they start radiating at
$\nu$; this time is calculated in Section 2. Electron distribution function
in the neighborhood of the desired LF is assumed to grow with radius
as $\propto R^x$, and the distribution attains its asymptotic power-law shape
at some radius $R_f\sim 2 R_0$. The rise of the light curve depends on 
$R_0$, $R_f$, $x$ and the deceleration radius; the rise also depends weakly on 
the type of CSM and the energy spectral index.

We show in Fig. 1 light curves for two different regimes $R_0<R_d<R_f$ (case 1) and 
$R_d < R_0 < R_f$ (case 2); for $R_0<R_f<R_d$ the light curve is similar to case 1  
except that between $t_f$ (the observer time corresponding to radius $R_f$) and $t_d$ the 
light curve rises as $\sim t^3$. Guided by the estimates provided in \S2 we
take $R_0 \sim R_d/3$ for 100 MeV photons, whereas for 10 GeV $R_0\sim R_d$.
We show the results for these choices of parameters in Fig. 1; note the
very steep rise of light curves which appear similar to the fast rise of the
observed 100 MeV light curve for GRB080916C as reported in Abdo et al. (2009a).
 
%%%%%%%%%%%%%%%%%%%%%%%%%%%%%%%%%Figure%%%%%%%%%%%%%%%%%%%%%%%%%%%%%%%%%%%%%

\begin{figure}
\begin{center}
\includegraphics[width=6.5cm, angle=0]{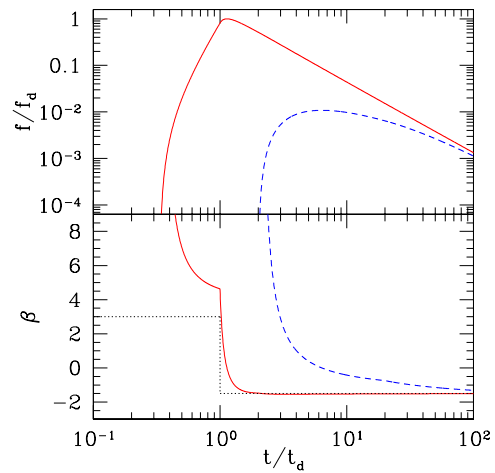}
\end{center}
\caption{{\it Top}. The expected external forward shock light curve
when the non-zero acceleration time of the emitting electrons is taken into account.
We plot the specific flux (normalized to the flux at $t_d$, $f_d$) versus 
time (normalized to the deceleration time, $t_d$).  We show two cases:
(1) $R_0<R_d<R_f$ (red solid line), and (2) $R_d<R_0<R_f$ (blue dashed line).
{\it Bottom}.  The light curve temporal slope, 
$\beta= d \ln(f/f_d)/ d \ln(t/t_d)$.  The horizontal black dotted line
shows the asymptotic value of the temporal decay index if we take 
electrons to accelerate instantaneously; for $t<t_d$ the light curve 
would rise as $t^3$.  However, we find that the external 
forward shock light curve rises faster than $t^3$ due to the finite time
it takes electrons to accelerate. }
%\label{fig01} % optional figure label, must be unique
\end{figure}

%%%%%%%%%%%%%%%%%%%%%%%%%%%%%%End Figure%%%%%%%%%%%%%%%%%%%%%%%%%%%%%%%%%%%%

\section{Conclusions}

In this paper we have investigated the acceleration
of electrons via diffusion shock acceleration 
in the external forward shock of GRBs, and its 
implications for the high-energy photon detection
by the {\it Fermi} Satellite. The external shock model,
with a weak magnetic field, has been proposed
as the origin of the observed $>$100 MeV emission detected by
the {\it Fermi} Satellite from a number of GRBs (Kumar \& Barniol Duran
2009, 2010). We find that high-energy electrons of Lorentz factor
$\sim 10^8$, required for producing $\sim$10 GeV photons via the
synchrotron process, can indeed be accelerated in an external
shock that is moving through a CSM with a magnetic field of strength 
a few tens of $\mu$G; they remain confined to the shock front as long as 
the upstream magnetic field is $\gae 10 \mu$G.

We have also calculated the time it takes for electrons to be 
accelerated to a Lorentz factor $\sim10^7$ so that they can radiate synchrotron 
photons at $\sim$100 MeV. We find this acceleration time to be a few seconds 
in the observer frame; this calculation took into account radiation losses 
suffered during the acceleration process. 
This result offers a straightforward explanation as to why, for most {\it Fermi} 
GRBs, 100 MeV photons are not observed right at the trigger time but a little 
later. This also explains, why 100 MeV photons 
are observed before GeV radiation: it takes electrons radiating at 
GeV energies even longer time to accelerate. Taking this acceleration 
time into consideration while calculating high-energy light curves, 
we find that the light curve rises very rapidly -- much faster than 
it does for the external forward shock model with instantaneous
electron acceleration for which the flux rises as $t^3$ when the
CSM has uniform density (the $t^3$ rise reflects the increasing number 
of swept-up electrons  before the blast wave decelerates).

The detection of the first 100 MeV 
photons at some fraction of the deceleration time, the longer delays 
in the detection of higher energy photons\footnote{Note that this 
possible trend in the data goes in the opposite direction than 
in the prompt $\sim 1$ MeV emission, where higher energy photons 
arrive {\it earlier} than lower energy photons in long GRBs and 
there is no lag detected for short GRBs (Norris et al. 1986,
Norris \& Bonnell 2006).} 
and the fast rise of the 100 MeV light 
curve, follow the expectation of the external forward shock model 
when the finite time for electron acceleration is taken into account.  
Detection of synchrotron photons of different energies provides
an upper limit for the radiation flux produced in the reverse shock
heated GRB-jet. For instance, the peak flux for the external reverse
shock emission --- if the peak of the spectrum is at a few eV ---
couldn't have been larger than about 300 mJy close to the deceleration time,
for GRB080916C, otherwise it would prevent electrons from accelerating
to a Lorentz factor of $\sim10^7$ so that they can produce synchrotron
photons of 100 MeV energy at early times (see Table 2). Similarly, 
the reverse-shock flux should be $\lae$ 20 mJy for GRB 080916C in
order that electrons in the forward shock are accelerated to a LF so
that they produce 1 GeV photons.

We speculate that the lack of $>$100MeV emission during the prompt 
phase of GRBs might be due to the presence of a bright optical source 
with observed flux larger than about 100 mJy,
which would prevent electrons from reaching high Lorentz factors.  This, 
coupled with the fact that GRBs with the largest LFs, which have small
 deceleration time, are the most likely bursts to be detected by {\it Fermi}
(Kumar \& Barniol Duran 2009) might explain the detection/non-detection 
 of $>100$MeV radiation from GRBs.

We note that the shock-compressed magnetic field scenario  
requires some cross-field diffusion of particles - presumably 
generated by turbulence - to allow them 
to travel back to the upstream (e.g. Achterberg et al. 2001, Lemoine et al. 2006). 
This turbulent layer probably occupies a small fraction 
of the downstream region as suggested by recent simulations by Sironi \& Spitkovsky (2010).
Therefore, the picture that seems to emerge from numerical simulations 
and {\it Fermi} observations, is that there might be a small region of turbulence 
behind the shock front that aids in the acceleration of particles across the shock, 
but that the radiation is mainly produced by particles that are swept downstream where 
the value of the downstream field is consistent with simple shock-compression of 
upstream field.

There exists also the possibility that the CSM seed field is actually  
a few $\mu$G and some instability produced ahead of the shock 
amplifies it to the value of a few tens of $\mu$G we infer by 
our modeling of {\it Fermi} GRBs (Kumar \& Barniol Duran 2009, 2010).
These instabilities have been studied by,  
e.g., Milosavljevi\'c \& Nakar 2006, Sironi \& Goodman 2007, 
Goodman \& MacFadyen 2008, Couch et al. 2008.
However, this {\it possible} amplification of a factor of $\sim$10
is much smaller than the amplification customarily invoked 
to explain afterglow observations.

We received a preprint from Piran \& Nakar (2010) soon after this paper was
completed. They have also considered the acceleration of electrons in 
the external shock.
% The main difference in our results, as far as we can tell, is that Piran \& Nakar 
%suggest a factor few larger magnetic field in the CSM than we find; this difference 
%arises because the synchrotron cooling frequency in their calculation is larger than 
%us by a factor $\sim10$.  

\section*{Acknowledgments}
RBD dedicates this work to the memory of Luis Matanzo, and thanks 
Jessa Barniol for her support during the writing of this manuscript.
We thank the anonymous referee for a constructive report and also thank
Tsvi Piran and Ehud Nakar for useful discussions.
This work has been funded in part by NSF grant ast-0909110.

%%%%%%%%%%%%%%%%%%%%Begin the Reference%%%%%%%%%%%%%%%%%%%%%%%%%%

%%%%%%%%%%%%%%%%%%%%End the Reference%%%%%%%%%%%%%%%%%%%%%%%%%%%

\end{document}